\documentclass[aps,twocolumn]{revtex4-1} 
\usepackage{graphicx}
\usepackage{epstopdf}
\usepackage{amsfonts,amsthm}
\usepackage{amsmath}
\usepackage{amssymb}
\usepackage{bm}
\def\vec#1{\boldsymbol #1}

\begin{document}
\title{Semi-Quantized Spin Pumping and Spin-Orbit Torques in Topological Dirac
Semimetals}

\author{
Takahiro Misawa$^1$ and Kentaro Nomura$^{2,3}$
}

\affiliation{$^1$Institute for Solid State Physics,~University of Tokyo,~5-1-5 Kashiwanoha, Kashiwa, Chiba 277-8581, Japan}
\affiliation{$^2$Institute for Materials Research,~Tohoku University,~Sendai 980-8577,~Japan}
\affiliation{$^3$Center for Spintronics Research Network, Tohoku University, Sendai 980-8577, Japan}


\begin{abstract}
We study the time-development processes of spin and charge transport
phenomena in a topological Dirac semimetal attached to a ferromagnetic
insulator with a precessing magnetization. Compared to conventional
normal metals, topological Dirac semimetals manifest a large inverse
spin Hall effect when a spin current is pumped from the attached
ferromagnetic insulator. It is shown that the induced charge current is
semi-quantized, i.e., it depends only on the distance between the two
Dirac points in momentum space and hardly depends on the disorder
strength when the system remains in the topological Dirac semimetal
phase. As an inverse effect, we show that the electric field applied to
the topological Dirac semimetal exerts a spin torque on the local
magnetization in the ferromagnetic insulator via the exchange
interaction and the semi-quantized spin Hall effect. Our study
demonstrates that the topological Dirac semimetal offers a
less-dissipative platform for spin-charge conversion and spin switching.
\end{abstract}

\flushbottom
\maketitle
%
%
\thispagestyle{empty}


\section*{Introduction}
Manipulation of magnetization direction by applying electric currents is
intended to be used in future magnetic devices, allowing information to
be written electrically~\cite{spintronics_review,spintronics_review1,spintronics_review2}.
Spin-orbit torques such as those induced by the spin-Hall effect~\cite{murakami2003,Sinova2015} and the Rashba-Edelstein
effect~\cite{rashba1984,edelstein1990} have recently been examined for a variety of
materials. 
{However, such current driven magnetization switching suffers from Joule heating problems for device applications.}
A significant effort has been made in the
search for materials having high efficiency~\cite{spintronics_review,spintronics_review1,spintronics_review2,Spin-current-book}. 
Obviously, one can expect a large spin-torque effect in systems with a strong
spin-orbit interaction~\cite{brataas2012}. However, theoretical studies on the
spin-torque effect in such strongly coupled spin-orbit systems are
beyond the scope of the conventional theory~\cite{tserkovnyak2005}.

As an inverse effect of the electrically induced spin-torque effect, a
charge current is generated in a metal by the precessing magnetization
of an attached ferromagnetic insulator~\cite{saitoh2006}. This phenomenon can
be interpreted as a combination of the spin-pumping effect and the
inverse spin-Hall effect or the inverse Rashba-Edelstein effect~\cite{sanchez2013,shen2014}. 
The strength of the spin-Hall effect and the inverse
spin-Hall effect is characterized by the spin-Hall angle $\theta_{\rm SH} = (2{e}/{\hbar}){j_{s}}/{j_{c}}$,
where {$j_{c}$} is the charge current generated by an
applied electric field and {$j_{s}$} is the spin current
induced by the spin-Hall effect. For typical metals {such as Pt, Au, and Ta}, the value of the
spin-Hall angle is $\sim$ 0.1~\cite{Spin-current-book,spin_torque_review}. Materials with larger spin-Hall
angles have been researched for application in devices.

Quantum spin Hall insulators~\cite{hasan2010,qi2011} have a quantized spin Hall
conductivity and a vanishing longitudinal conductivity. In systems with
boundaries the spin-momentum locked gapless edge states lie inside the
bulk gap. Several theoretical studies have been conducted on coupled
spin dynamics and charge transport at the interface between a quantum
spin Hall insulator and a ferromagnetic material, such as magnetically
generated charge currents~\cite{Qi_2008,mahfouzi2010,chen2010,luo2016,soleimani2017,meng2014,dora2012,vajna2016}. However, magnetization
reversal in these systems might be difficult because the interface area
is small. To exert a large spin torque on the magnetization and reverse
its direction, a two-dimensional interface is necessary. Recently, the
interface between a three-dimensional (3D) topological insulator and a
ferromagnetic material has been realized experimentally. Relatively
large spin Hall angles measured using spin transfer torque ferromagnetic
resonance, spin-charge conversion, and magnetization reversal have been
reported~\cite{mellnik2014,Shiomi2014,fan2014,kondou2016,jamali2015}. With the successful research on the
spintronics phenomena using topological insulators, studies on a wider
range of 3D topological materials with higher functionalities are
desired.

In this work, we consider a topological Dirac semimetal (TDSM)~\cite{wang2012_Na3Bi,morimoto2014,yang2014,ZKLiu2014Na3Bi,wang2013_Cd3As2,
ZKLiu2014Cd3As2,Neupane2014,borisenko2014} as an effective platform for spin-charge conversion and
study magnetically induced charge pumping and current-induced
magnetization reversal. TDSMs are 3D gapless materials with pair(s) of
doubly degenerate Dirac cones, separated in the momentum space along a
rotational axis (Fig. 1 {(a)}) and protected by rotational
symmetry~\cite{morimoto2014,yang2014}. The degeneracy is attributed to time-reversal
and space-inversion symmetries. Na\textsubscript{3}Bi~\cite{wang2012_Na3Bi,ZKLiu2014Na3Bi}
and Cd$_{3}$As$_{2}$~\cite{wang2013_Cd3As2,ZKLiu2014Cd3As2,Neupane2014,borisenko2014} have been
theoretically and experimentally confirmed to be TDSMs. One of the
prominent features of this system, which plays an essential role in this
work, is that when the Fermi level resides at the Dirac point, the
longitudinal conductivity vanishes (i.e., $\sigma_{xx} = 0$)~\cite{fradkin1986critical,ominato2014quantum} while the spin Hall conductivity is semi-quantized in the
bulk limit (i.e., 
$\sigma^{z}_{xy} =\frac{e}{4\pi^{2}}\Delta k$, 
where $\Delta k$ is the distance
between Dirac points)~\cite{burkov2016}, {indicating that the spin Hall angle
${\theta}_{\rm SH}$ diverges in the bulk.} The large spin Hall angle causes
the strong spin-torque effect induced by the small longitudinal charge
current. Another important point is that helical surface modes emerge at
the boundary of a TDSM in which the spin-up electrons go one way while
the spin-down electrons go the opposite way. Under an applied electric
field, a charge current flows with transverse spin polarization similar
to the Rashba-Edelstein effect. Naively, one can expect that the spin
polarization generated in the helical modes in TDSMs is larger than that
in the Rashba systems and the surface of topological insulators
{since only $z$ component of spin is approximately conserved}.

Using the numerical time-development formalism of quantum systems~\cite{suzuki1994,nakanishi1997}, we study pumped charge currents by precessing
magnetization and spin-orbit torque. This formalism enables us to
analyze the transport phenomena and spin-torque effects in the
heterostructures beyond conventional theories such as the linear
response theory. We claim that the Dirac electrons exert spin-orbit
torques on the local magnetization of the ferromagnet at the interface,
even when the Fermi level resides at the Dirac point. As in conventional
bilayer systems of a ferromagnet and a metal, anti-dumping torques are
induced by the bulk spin Hall current and a field-like torque is induced
by the accumulated electron spins at the interface.


\section*{Results}

\subsection*{Model and Methods}
TDSMs are characterized by a pair of Dirac points along the
${k}_{z}$ axis, stabilized by discrete rotational
symmetry (see Fig. 1 {(a)}). 
To describe the electronic states in
a TDSM on a simple lattice system, 
we employ the effective tight-binding Hamiltonian
\begin{align}
&{\cal H}_{\rm TDSM}(\vec{k})=\sin{k_{x}}\alpha_{3}+\sin{k_{y}}\alpha_{4}+ \notag \\
&(2+m-\cos{k_{x}}-\cos{k_{y}}-\cos{k_{z}})\alpha_{5},
\end{align}
which is connected, in the long wavelength limit, to the four-band
${k}\cdot{p}$ Hamiltonian derived for Na$_{3}$Bi~\cite{wang2013_Cd3As2}, where
\begin{align}
\alpha_{3}=
\begin{pmatrix}
0          &\  \sigma_{z} \\
\ \sigma_{z} & 0 \\
\end{pmatrix},\ 
\alpha_{4}=
\begin{pmatrix}
0 ~& \ -{\rm i} \\
 {\rm i} ~& \ \ 0 \\
\end{pmatrix}, \ 
\alpha_{5}=
\begin{pmatrix}
I ~~& 0 \\
0 ~~& -I \\
\end{pmatrix}
\end{align}
are $4\times 4$ matrices, $\sigma_i$ $({i} = {x,y,z})$
being the Pauli matrices acting on the spin indices. The Dirac points
are located at ${\vec{k}}_{\rm D} = (0,0,\pm\cos^{-1}{(m)})$ 
for ${|m|}<1$. 
We consider a TDSM attached to a ferromagnetic insulator
at ${x} = 0$ plane (see Fig. 1~{(b)}), whose Hamiltonian is
given by ${H}(\emph{t}) = {H}_{\rm TDSM} + {H}_{\rm exc}({t})$, where
\begin{align}
H_{\rm TDSM}
&=\sum_{j,\mu=x,y,z}
\Big(c_{j+\vec{e}_{\mu}}^{\dagger}T_{\mu}c_{j}+{\rm H.c.}\Big)\notag \\
&+(2+m)\sum_{j}c_{j}^{\dagger}\alpha_{5}c_{j}
 \label{eq:TDSM}
\end{align}
is the real space representation in the second quantization formalism.
In Eq. (\ref{eq:TDSM}), ${c}^{\dagger}_{j}({c}_{j})$ represents the four-component fermion
creation (annihilation) operator defined on a site ${j}$ on the 3D
cubic lattice spanned by three orthogonal unit vectors
$\vec{e}_{\mu}={x,y,z}$. The matrix is defined as
$T_{x}=(-\alpha_{5}+i\alpha_{3})/2$,
$T_{y}=(-\alpha_{5}+i\alpha_{4})/2$,
$T_{z}=\alpha_{5}/2$.
On the other hand,
\begin{equation}
H_{\rm exc}(t)= \sum_j J\hat{\vec{M}}(t)\cdot(J_{0}(x)c_{j}^{\dagger}\vec{s}c_{j}+J_{3}(x)c^{\dagger}_j(\vec{s}\alpha_{5})c_j),
\label{eq:exchange}
\end{equation}
describes the exchange interaction at the interface between the TDSM and the ferromagnet. 
$\hat{\vec{M}}$ is the unit vector
in the direction of local magnetization in the ferromagnetic insulator,
${J}_{0}(x)=J_0\times\exp(-x/\xi)$ and 
${J}_{3}({x})=J_{3}\times\exp(-x/\xi)$ 
are the coupling
constants of the exchange interactions with the penetration length $\xi$,
$\vec{s} = {\rm diag}(\vec{\sigma},\vec{\sigma})$ 
is the spin operator of the electrons in the TDSM. For simplicity, we take
${J}_{0}={J}_{3}={J}$ in this study. 
We perform the calculation for a 3D system whose size is given by ${L_{x}}\times{L}_{y}\times{L}_{z}$.
{
In this paper, we only treat the TDSM protected by the spin 
Chern number. We note that a different type
of the TDSM protected by the mirror Chern number is proposed~\cite{Chang_TDSM}.
} 

First, we consider the spin and charge pumping due to a precessing
magnetization. The ferromagnetic insulator is assumed to be excited by
microwave radiation (as done in the experiment~\cite{saitoh2006}) resulting in
a steady precession about the effective field close to the ferromagnetic
resonance condition:
\begin{align}
\hat{\vec{M}}(t)=(\sqrt{1-M_{0}^2}\cos(\omega t),\sqrt{1-M_{0}^2}\sin(\omega t),M_{0}),
\label{eq:M}
\end{align}
in Eq. (\ref{eq:M}), where $\omega = 2\pi/T$ is the frequency with ${T}$ being
the period of the precession. In the following analysis, we set ${T}=20$
and ${M}_{0} = 0$. When the magnetization begins to
precess, an electronic state of the TDSM evolves to a non-equilibrium
state $\Psi(t)$. $\Psi({t})$ is obtained by solving the time-dependent Shr\"{o}dinger equation
$i(\partial/\partial t)|\Psi(t)\rangle=H(t)|\Psi(t)\rangle$
numerically with a time step of $\Delta{t}$. The time development of the
wave function is given by $\Psi({t}+\Delta {t})={U}({t}+\Delta{t},{t})\Psi({t})$, 
where ${U}({t} + \Delta{t}$, $U(t)$ is the unitary time-evolution
operator defined as
\begin{align}
U(t+\Delta t,t)=\mathcal{T}\exp\Big[-{\rm i}\int_{t}^{t+\Delta t}dt'H(t')\Big].
\end{align}
Here, $\mathcal{T}$ is the time-ordering operator. Using the formula
given in literature~\cite{suzuki1994,nakanishi1997}, we decompose ${U}({t} + \Delta{t},{t})$ 
into a product of small exponential operators and
perform real-time evolution as matrix-vector multiplication (for
details, see the 
Methods section). In this method, the
diagonalization of the Hamiltonian is necessary only for preparing the
initial wave function and the numerical cost is significantly reduced.
In the following analysis, we impose a periodic boundary condition in
the ${z}$ direction and a fixed boundary condition in the ${x}$
direction. Depending on the quantities to be computed, the periodic
boundary condition or the fixed boundary condition is implemented in the ${y}$ direction.
{In the following, we set the Fermi energy at the Dirac point.}

\begin{figure}[ht]
\centering
\includegraphics[width=\linewidth]{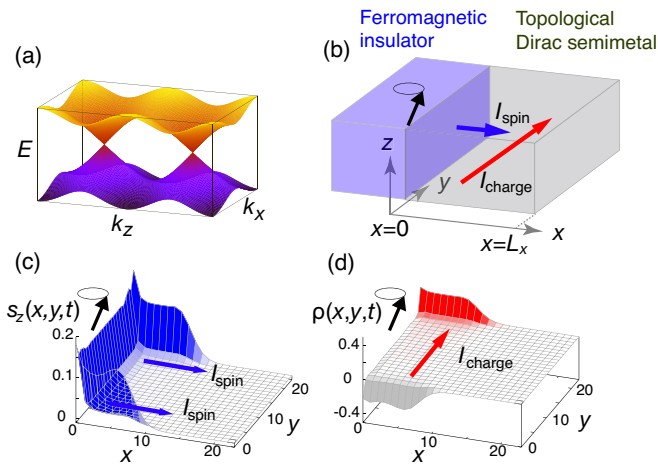}
\caption{
Setup of a topological Dirac semimetal attached to a ferromagnetic insulator and snapshot of charge and spin density. 
(a) TDSM has a couple of Dirac points where the two bands touch at the points which reside on the $k_z$ axis in the momentum space. 
(b) Schematic illustration of the bilayer system of a ferromagnetic insulator and a TDSM. 
Precessing magnetization causes spin injection and generates transverse charge current. 
(c) Snapshot of the $z$ component of spin polarization $s_z(x,y,t)$ at $t/T = 1.0$ for $L_x = L_y = 2L_z = 24$, 
$J = 0.5$,$T=20$,$\xi=0.1$, and $m = 0$. 
The spin polarization is injected into the TDSM.
(d) Snapshot of the charge polarization $\rho(x,y,t)$ at $t/T = 1.0$. 
The charge polarization is induced by the injected spins.
}
\label{fig:F1}
\end{figure}

\subsection*{Spin density and charge density}
To characterize the time evolution of the spin and charge densities in
the TDSM, we introduce the quantity 
${n}_{\sigma}(x,y,z)|_t=\langle\Psi(t)|c_{j\sigma}^{\dagger}c_{j\sigma}|\Psi(t)\rangle $
as the density of electrons with spin $\sigma=\uparrow,\downarrow$ 
at time ${t}$ and position 
${j} = ({x},{y},{z})$. The precession motion of the magnetization
starts at ${t} = 0$. 
The subsequent time development of the electrons is given by

\begin{eqnarray}
  N_{\sigma}(x,y,t)\equiv \frac{1}{L_z}\sum_z\Big[n_{\sigma}(x,y,z)\big|_t- n_{\sigma}(x,y,z)\big|_{t=0}\Big].
 \nonumber\\
\end{eqnarray}

Since we are interested in the spin and charge propagating in the
${x}$ and/or $y$ directions, the unimportant variable ${z}$
is integrated. For this purpose, we terminate the system by applying the
fixed boundary condition in the ${x}$ and ${y}$ directions, while
the periodic boundary condition is implemented in the ${z}$
direction. The time evolution of the charge and spin densities is given
by
\begin{eqnarray}
 \rho(x,y,t) &=& -e(N_{\uparrow}(x,y,t) + N_{\downarrow}(x,y,t))  \notag \\
 s_z(x,y,t) &=& \   \frac{1}{2}(N_{\uparrow}(x,y,t) - N_{\downarrow}(x,y,t)). \notag 
\end{eqnarray}

In Figs. 1 {(c)} and {(d)}, we present the snapshots of
$\rho(x,y,t)$ and $s_{z}(x,y,t)$
at ${t/T} = 1$. Immediately after the
magnetization begins to precess, spin polarization of the Dirac
electrons is generated near the interface at ${x} = 0$. The
accumulated spins then begin to propagate into the TDSM. As the TDSM
possesses the gapless surface states while the density of states
vanishes in the bulk, the spins propagate mainly on the surface
($y = 0$ and $y = L_{y}-1$) as shown in Fig.~1{(c)}. 
In response to the spin propagation, the electrons are
pumped in the ${y}$ direction as shown in Fig. 1 {(d)}. This
development of the charge polarization is related to the charge current
flowing in the ${y}$ direction, anticipated from 
{from the inverse spin Hall effect and the inverse Rashba-Edelstein effect~\cite{Ganichev_Nature2002,Shen_PRL2014}}.

\subsection*{Charge current}
Next, we directly compute the charge current induced by the precessing
magnetization. For this purpose, we apply the periodic boundary
condition in the ${y}$ direction (the direction in which the current
flows), in addition to the ${z}$ direction. The charge current
operator is given by

\[\begin{matrix}
I_{c}^{y} & = & - {\rm i}\sum_{j}^{}(c_{j + \vec{e}_{y}}^{\dagger}T_{y}c_{j} - c_{j}^{\dagger}T_{y}^{\dagger}c_{j +  \vec{e}_{y}}).\ \ (9) \\
\end{matrix}\]

Its expectation value
$\langle\Psi(t)| I_c^y |\Psi(t)\rangle$ 
is plotted as a function of time in Fig. 2 {(a)}. After the
magnetization begins to precess, the charge current increases at the
initial stage of the time evolution. For ${t} \ge 2{T}$, the
charge current converges to a constant value with some oscillations. We
obtain the average value of the charge current from the relation:
\begin{equation}
 \bar{I}_c= \frac{1}{\Delta t}\int_{t_0}^{t_0+\Delta t}\!dt  \langle\Psi(t)| I_c^y |\Psi(t)\rangle
\end{equation}
We consider ${t}_{0} = 2{T}$ and $\Delta {t}={T}$. 
Figure 2 {(b)} shows the $J$-dependence of
(${\bar{I}}_{c}$); the figure clearly indicates that
(${\bar{I}}_{c}$) increases with ${J}$ from zero and
converges to a constant value when ${J}$ is large enough. We found
that the average value of the current is given by
\begin{equation}
  \bar{I}_c=\nu\frac{e}{2\pi}\frac{\partial \phi}{\partial t},
\label{quantized_current}
\end{equation}
where 
{$\nu=L_z\Delta k/2\pi$ is the number of the helical channels at the surface and}
$\phi={\rm tan}^{-1}(M_y/M_x)$.
This is examined by studying the dependence of the distance between the
Dirac points in momentum space, 
$\Delta k = 2|\vec{k}_{\rm D}|= 2\cos^{-1}(m)$,
on the charge current by tuning the
parameter $m$. By increasing $\Delta k$, we observe that the pumped
charge current increases linearly with $\Delta {k}$ as shown in Fig. 2 {(c)}. 
This result indicates that the charge current is governed
by Fermi arcs in the TDSM since the Fermi arcs increase in length as
$\Delta k$ increases.

\begin{figure}[h!]
\centering
\includegraphics[width=\linewidth]{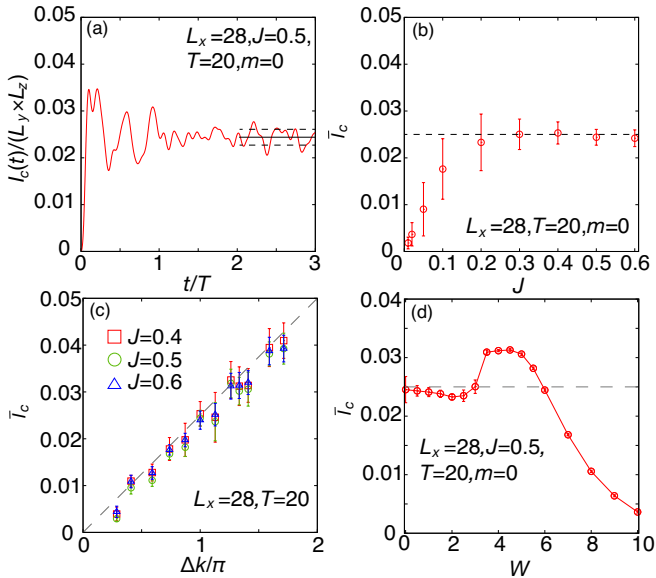}
\caption{
Charge current as a function of time, exchange coupling, distance between Dirac points, and disorder strength.
(a) Time dependence of the charge current for the case $J = 0.5$,$T = 20$, $m = 0$, $\xi = 0.1$, and 
$L_x = 2L_y = 2L_z = 28$. We regard the width of the oscillation (standard deviation) of the charge current in 
$3\ge t/T \ge 2$ as the error bars (shown in broken lines). 
(b) Averaged charge current ($\bar{I}_c$ ) as a function of the exchange coupling $J$. 
(c) Averaged charge current  ($\bar{I}_c$ ) as a function of the distance between the Dirac points for $J = 0.4,0.5$, and $0.6$ ($L_x = 28$). 
(d) Averaged charge current   ($\bar{I}_c$ ) as a function of the disorder strength $W$. 
}
\label{fig:F2}
\end{figure}

\subsection*{Effects of disorder}
We consider the effects of disorder by introducing site-dependent random
potentials
\begin{align}
V=\sum_{i}\epsilon_{i}c_{i}^{\dagger}c_{i}, \notag
\end{align}
where $\epsilon_{i}\in [-W/2:W/2]$ is random
number and ${W}$ characterizes the strength of disorder. Figure 2
{(d)} shows the disorder dependence of the induced charge
current. We take 72 independent realizations for performing the disorder
average. We calculate both the standard errors of the current with
respect to the disorders and the disorder-average oscillations of the
current, and regard the larger one as the error bars. At weak disorder
($0\lesssim W\lesssim 3$), the charge current is insensitive to the disorder
strength and approximately takes the semi-quantized value. 
At strong disorder ($3\lesssim\emph{W}$) where the TDSM
phase is broken by disorder, the time-averaged value of the induced
charge current deviates from the semi-quantized value: it first
increases and then decreases as ${W}$ increases. 
These results indicate that the magnetically induced charge current is robust against
disorder when the system is in the TDSM phase. The increase in charge
current in the strong disorder regime could be explained as follows. It
is known that the density of states and the bulk conductivity at the
Dirac point remains vanishing in the Dirac semimetal phase~\cite{fradkin1986critical}.
When the disorder strength exceeds the critical value, the system turns
into the diffusive metallic phase where the density of states and the
bulk conductivity become finite. The increased density of states could
enhance the bulk contribution to the magnetically induced charge current.
{We note that the similar disorder induced phase transition 
occurs for the Weyl semimetals~\cite{Chen_PRL2015,Shapourian_PRB2016}.}

\subsection*{Magnetization switching and spin torque}
From the above discussion, one can infer that the magnetization dynamics
in the ferromagnetic insulator generates the charge current in the TDSM.
In the rest of this work, we study the spin-torque effect and dynamics
of the magnetization induced by the electric field in the TDSM.
Magnetization dynamics in the ferromagnetic insulator are described by
the phenomenological Landau-Lifshitz-Gilbert (LLG) equation:

\begin{align}
&\frac{d\hat{\vec{M}}}{{dt}} = - \gamma\hat{\vec{M}} \times \vec{H}_{\text{eff}} + \alpha\hat{\vec{M}} \times 
\frac{d\hat{\vec{M}}}{{dt}} \notag\\
&+{\frac{1}{\hbar}}\hat{\vec{M}} \times (\sum_{x}^{}J_{0}(x)\langle\vec{s}\rangle_{x} 
+ J_{3}(x)\langle\vec{s}\alpha_{5}\rangle_{x})
\label{eq:LLG}
\end{align}

Here, we assume that the magnetic insulator is thin enough and the
magnetization is uniform in space. In Eq. (\ref{eq:LLG}), 
$\gamma$ is the gyromagnetic ratio, 
$\alpha$ is the Gilbert damping constant,
$\langle \vec{s}\rangle_{x}$ is the spin density of
itinerant Dirac electrons at $x$, and
$\vec{H}_{\rm eff} = K_{z}(M_{z}\vec{e}_{z})$
is the effective magnetic field, which induces the easy-axis anisotropy.
The spin density of the Dirac electrons at time ${t}$ and position
${j}=({x},{y},{z})$ is given by
$\vec{s}(x,y,z)|_t=\sum_{\sigma,\sigma'}  \langle\Psi(t)|c_{\sigma j}^{\dag}\vec{\sigma}_{\sigma\sigma'}^{}c_{\sigma' j}^{}|\Psi(t)\rangle$.
The averaged spin density at ${x}$ corresponds to
 $\langle \vec{s}\rangle_{x,t}=(1/A)\sum_{y,z}\vec{s}(x,y,z)|_t$,
${A}$ being the area of the interface. 
The time evolution of $|\Psi(t)\rangle$ and $\hat{\vec{M}}$ is obtained by
simultaneously solving the time-dependent Schrödinger equation and the
LLG equation. A typical dynamical behavior of the magnetization is shown
in Fig. 3, where an electric field is turned on at $t = 0$ via the
Peierls substitution of a time-dependent vector potential.
Here, we take $\gamma=-1$, $\alpha = 0.1$, $J = 0.5$,
$L_{x} = 4L_{y} = 4L_{z} = 32$, ${W}=0.5$,
${K}_{z}=-0.01$, $\xi=0.1$ and $E_{y} = \frac{\Phi_{0}}{eL_{y}T}$, $\Phi_{0}$ being the flux
quantum. We take 50 independent realizations for performing the disorder
average. The direction of the magnetization, originally oriented in $-z$ direction
($\vec{M}=(0,(1-{M}_{0}^{2})^{1/2},{M}_{0}),{M}_{0}=0.99$)
starts to change due to the electrically induced spin torque. After a
sufficient passage of time, the magnetization is reversed to $+z$
direction.

Due to the spin Hall effect, when an electric field is applied in the
${y}$ direction, a spin current ${I_{z}}$ is
generated and flows in the ${x}$ direction (Fig. 3 {(a)}). At
the interface ${x} = 0$, the finite spin current is transferred to a
spin torque acting on the local magnetization in the ferromagnet.
When the spin-orbit interaction at the interface is negligible, this
spin-Hall-current induced torque is expressed as the damping-like (DL)
torque, 
$\vec{T}_{\text{DL}}\propto I^z_{\rm spin} \hat{\vec{M}}\times(\hat{\vec{z}}\times\hat{\vec{M}})$.
{On the other hand, the contribution of the nonequilibrium spin density 
at the helical surface state of the TDSM~\cite{Mellnik_2014nature} to the torque 
is known as the field-like (FL) torque},
$\vec{T}_{\text{FL}}\propto S^z_{\rm edge}\hat{\vec{z}}\times\hat{\vec{M}}$, where
${S}^{z}_{\rm edge}$ is the accumulated spin
density at the interface. Figure 3 {(b)} indicates that the
damping-like torque and the field-like torque have the same order of
magnitude because both $M_{x}$,
$M_{y}$ and ${M}_{z}$ become finite at
the initial stage of the magnetization switching.

\begin{figure}[ht]
\centering
\includegraphics[width=\linewidth]{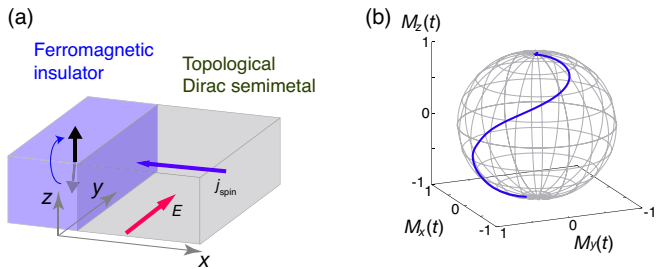}
\caption{
Set up of spin switching and typical trajectory of magnetization. 
(a)~Schematic illustration for magnetization switching. The electric field is applied in the $y$ direction and the spin current flows in the $x$ direction. 
(b)~Typical dynamical behavior of the local magnetization caused by the spin torque at the interface. 
At $t=0$, an electric field is switched on. The induced spin current exerts a spin torque on the local magnetization in the ferromagnetic insulator. 
}
\label{fig:F3}
\end{figure}

\section*{Discussion}
In this work, we studied dynamically injected spins into the TDSM from
the ferromagnetic insulator with precessing magnetization. The spins
propagate and generate the transverse charge current in the TDSM. The response
is semi-quantized in the sense that the charge current depends only on the
distance between two Dirac points in the momentum space and is robust
against disorder when the system remains in the TDSM phase. Moreover, we
have studied also its inverse response: when an electric field is
applied to the TDSM, a spin torque is exerted on the local magnetization
in the ferromagnetic insulator and the direction is reverted. 
The induced damping-like torque and field-like torque are in same order of magnitude. 
We note that the magnitudes of the spin Hall conductivity
$\sigma_{xy}^{z}$ for Na$_{3}$Bi~\cite{wang2012_Na3Bi} and
Cd$_{3}$As$_{2}$~\cite{wang2013_Cd3As2} are comparable with
those of conventional metals such as {$\beta$-Ta~\cite{Spin-current-book,Liu_Science2012,Qiao_PRB2018}}. 
Since the density of states at the Dirac point is vanishingly small, we expect
a large spin torque for the TDSM, while the Joule heating accompanying
the generation of spin torque is extremely lower than that in
conventional metals~\cite{Liu_Science2012}.

\section*{Acknowledgements}
We thank Y. Araki, G. Bauer, K-I. Imura, and K. Kobayashi for useful discussions.
Our calculation was partly carried out at the Supercomputer Center, Institute for Solid State Physics, University of Tokyo. 
KN was supported by JSPS KAKENHI Grants No. JP15H05854 and No. JP17K05485, and JST CREST Grant No. JPMJCR18T2. 
TM was supported by JSPS KAKENHI Grants No. JP16H06345, No. JP16K17746, No. JP19K03739, and by Building of Consortia for the Development of Human Resources 
in Science and Technology from the MEXT of Japan. 


\begin{thebibliography}{50}
\expandafter\ifx\csname natexlab\endcsname\relax\def\natexlab#1{#1}\fi
\expandafter\ifx\csname bibnamefont\endcsname\relax
  \def\bibnamefont#1{#1}\fi
\expandafter\ifx\csname bibfnamefont\endcsname\relax
  \def\bibfnamefont#1{#1}\fi
\expandafter\ifx\csname citenamefont\endcsname\relax
  \def\citenamefont#1{#1}\fi
\expandafter\ifx\csname url\endcsname\relax
  \def\url#1{\texttt{#1}}\fi
\expandafter\ifx\csname urlprefix\endcsname\relax\def\urlprefix{URL }\fi
\providecommand{\bibinfo}[2]{#2}
\providecommand{\eprint}[2][]{\url{#2}}

\bibitem[{\citenamefont{Brataas and Hals}(2014)}]{spintronics_review}
\bibinfo{author}{\bibfnamefont{A.}~\bibnamefont{Brataas}} \bibnamefont{and}
  \bibinfo{author}{\bibfnamefont{K.~M.} \bibnamefont{Hals}},
  \bibinfo{journal}{Nature nanotechnology} \textbf{\bibinfo{volume}{9}},
  \bibinfo{pages}{86} (\bibinfo{year}{2014}).

\bibitem[{\citenamefont{Hoffmann and Bader}(2015)}]{spintronics_review1}
\bibinfo{author}{\bibfnamefont{A.}~\bibnamefont{Hoffmann}} \bibnamefont{and}
  \bibinfo{author}{\bibfnamefont{S.~D.} \bibnamefont{Bader}},
  \bibinfo{journal}{Physical Review Applied} \textbf{\bibinfo{volume}{4}},
  \bibinfo{pages}{047001} (\bibinfo{year}{2015}).

\bibitem[{\citenamefont{Kent and Worledge}(2015)}]{spintronics_review2}
\bibinfo{author}{\bibfnamefont{A.~D.} \bibnamefont{Kent}} \bibnamefont{and}
  \bibinfo{author}{\bibfnamefont{D.~C.} \bibnamefont{Worledge}},
  \bibinfo{journal}{Nature nanotechnology} \textbf{\bibinfo{volume}{10}},
  \bibinfo{pages}{187} (\bibinfo{year}{2015}).

\bibitem[{\citenamefont{Murakami et~al.}(2003)\citenamefont{Murakami, Nagaosa,
  and Zhang}}]{murakami2003}
\bibinfo{author}{\bibfnamefont{S.}~\bibnamefont{Murakami}},
  \bibinfo{author}{\bibfnamefont{N.}~\bibnamefont{Nagaosa}}, \bibnamefont{and}
  \bibinfo{author}{\bibfnamefont{S.-C.} \bibnamefont{Zhang}},
  \bibinfo{journal}{Science} \textbf{\bibinfo{volume}{301}},
  \bibinfo{pages}{1348} (\bibinfo{year}{2003}).

\bibitem[{\citenamefont{Sinova et~al.}(2015)\citenamefont{Sinova, Valenzuela,
  Wunderlich, Back, and Jungwirth}}]{Sinova2015}
\bibinfo{author}{\bibfnamefont{J.}~\bibnamefont{Sinova}},
  \bibinfo{author}{\bibfnamefont{S.~O.} \bibnamefont{Valenzuela}},
  \bibinfo{author}{\bibfnamefont{J.}~\bibnamefont{Wunderlich}},
  \bibinfo{author}{\bibfnamefont{C.~H.} \bibnamefont{Back}}, \bibnamefont{and}
  \bibinfo{author}{\bibfnamefont{T.}~\bibnamefont{Jungwirth}},
  \bibinfo{journal}{Rev. Mod. Phys.} \textbf{\bibinfo{volume}{87}},
  \bibinfo{pages}{1213} (\bibinfo{year}{2015}).

\bibitem[{\citenamefont{Bychkov and Rashba}(1984)}]{rashba1984}
\bibinfo{author}{\bibfnamefont{Y.~A.} \bibnamefont{Bychkov}} \bibnamefont{and}
  \bibinfo{author}{\bibfnamefont{{\'E}.~I.} \bibnamefont{Rashba}},
  \bibinfo{journal}{JETP lett} \textbf{\bibinfo{volume}{39}},
  \bibinfo{pages}{78} (\bibinfo{year}{1984}).

\bibitem[{\citenamefont{Edelstein}(1990)}]{edelstein1990}
\bibinfo{author}{\bibfnamefont{V.~M.} \bibnamefont{Edelstein}},
  \bibinfo{journal}{Solid State Communications} \textbf{\bibinfo{volume}{73}},
  \bibinfo{pages}{233} (\bibinfo{year}{1990}).

\bibitem[{\citenamefont{S.~Maekawa and Kimura}()}]{Spin-current-book}
\bibinfo{author}{\bibfnamefont{E.~S.} \bibnamefont{S.~Maekawa},
  \bibfnamefont{S.~O.~Valenzuela}} \bibnamefont{and}
  \bibinfo{author}{\bibfnamefont{T.}~\bibnamefont{Kimura}},
  \emph{\bibinfo{title}{Spin current}}, \bibinfo{howpublished}{(Oxford
  University Press, 2012)}.

\bibitem[{\citenamefont{Brataas et~al.}(2012)\citenamefont{Brataas, Kent, and
  Ohno}}]{brataas2012}
\bibinfo{author}{\bibfnamefont{A.}~\bibnamefont{Brataas}},
  \bibinfo{author}{\bibfnamefont{A.~D.} \bibnamefont{Kent}}, \bibnamefont{and}
  \bibinfo{author}{\bibfnamefont{H.}~\bibnamefont{Ohno}},
  \bibinfo{journal}{Nature materials} \textbf{\bibinfo{volume}{11}},
  \bibinfo{pages}{372} (\bibinfo{year}{2012}).

\bibitem[{\citenamefont{Tserkovnyak et~al.}(2005)\citenamefont{Tserkovnyak,
  Brataas, Bauer, and Halperin}}]{tserkovnyak2005}
\bibinfo{author}{\bibfnamefont{Y.}~\bibnamefont{Tserkovnyak}},
  \bibinfo{author}{\bibfnamefont{A.}~\bibnamefont{Brataas}},
  \bibinfo{author}{\bibfnamefont{G.~E.} \bibnamefont{Bauer}}, \bibnamefont{and}
  \bibinfo{author}{\bibfnamefont{B.~I.} \bibnamefont{Halperin}},
  \bibinfo{journal}{Reviews of Modern Physics} \textbf{\bibinfo{volume}{77}},
  \bibinfo{pages}{1375} (\bibinfo{year}{2005}).

\bibitem[{\citenamefont{Saitoh et~al.}(2006)\citenamefont{Saitoh, Ueda,
  Miyajima, and Tatara}}]{saitoh2006}
\bibinfo{author}{\bibfnamefont{E.}~\bibnamefont{Saitoh}},
  \bibinfo{author}{\bibfnamefont{M.}~\bibnamefont{Ueda}},
  \bibinfo{author}{\bibfnamefont{H.}~\bibnamefont{Miyajima}}, \bibnamefont{and}
  \bibinfo{author}{\bibfnamefont{G.}~\bibnamefont{Tatara}},
  \bibinfo{journal}{Applied physics letters} \textbf{\bibinfo{volume}{88}},
  \bibinfo{pages}{182509} (\bibinfo{year}{2006}).

\bibitem[{\citenamefont{S{\'a}nchez et~al.}(2013)\citenamefont{S{\'a}nchez,
  Vila, Desfonds, Gambarelli, Attan{\'e}, De~Teresa, Mag{\'e}n, and
  Fert}}]{sanchez2013}
\bibinfo{author}{\bibfnamefont{J.~R.} \bibnamefont{S{\'a}nchez}},
  \bibinfo{author}{\bibfnamefont{L.}~\bibnamefont{Vila}},
  \bibinfo{author}{\bibfnamefont{G.}~\bibnamefont{Desfonds}},
  \bibinfo{author}{\bibfnamefont{S.}~\bibnamefont{Gambarelli}},
  \bibinfo{author}{\bibfnamefont{J.}~\bibnamefont{Attan{\'e}}},
  \bibinfo{author}{\bibfnamefont{J.}~\bibnamefont{De~Teresa}},
  \bibinfo{author}{\bibfnamefont{C.}~\bibnamefont{Mag{\'e}n}},
  \bibnamefont{and} \bibinfo{author}{\bibfnamefont{A.}~\bibnamefont{Fert}},
  \bibinfo{journal}{Nature communications} \textbf{\bibinfo{volume}{4}},
  \bibinfo{pages}{2944} (\bibinfo{year}{2013}).

\bibitem[{\citenamefont{Shen et~al.}(2014{\natexlab{a}})\citenamefont{Shen,
  Vignale, and Raimondi}}]{shen2014}
\bibinfo{author}{\bibfnamefont{K.}~\bibnamefont{Shen}},
  \bibinfo{author}{\bibfnamefont{G.}~\bibnamefont{Vignale}}, \bibnamefont{and}
  \bibinfo{author}{\bibfnamefont{R.}~\bibnamefont{Raimondi}},
  \bibinfo{journal}{Phys. Rev. Lett.} \textbf{\bibinfo{volume}{112}},
  \bibinfo{pages}{096601} (\bibinfo{year}{2014}{\natexlab{a}}).

\bibitem[{\citenamefont{Li et~al.}(2019)\citenamefont{Li, Edmonds, Liu, Zheng,
  and Wang}}]{spin_torque_review}
\bibinfo{author}{\bibfnamefont{Y.}~\bibnamefont{Li}},
  \bibinfo{author}{\bibfnamefont{K.~W.} \bibnamefont{Edmonds}},
  \bibinfo{author}{\bibfnamefont{X.}~\bibnamefont{Liu}},
  \bibinfo{author}{\bibfnamefont{H.}~\bibnamefont{Zheng}}, \bibnamefont{and}
  \bibinfo{author}{\bibfnamefont{K.}~\bibnamefont{Wang}},
  \bibinfo{journal}{Advanced Quantum Technologies}
  \textbf{\bibinfo{volume}{2}}, \bibinfo{pages}{1800052}
  (\bibinfo{year}{2019}).

\bibitem[{\citenamefont{Hasan and Kane}(2010)}]{hasan2010}
\bibinfo{author}{\bibfnamefont{M.~Z.} \bibnamefont{Hasan}} \bibnamefont{and}
  \bibinfo{author}{\bibfnamefont{C.~L.} \bibnamefont{Kane}},
  \bibinfo{journal}{Reviews of modern physics} \textbf{\bibinfo{volume}{82}},
  \bibinfo{pages}{3045} (\bibinfo{year}{2010}).

\bibitem[{\citenamefont{Qi and Zhang}(2011)}]{qi2011}
\bibinfo{author}{\bibfnamefont{X.-L.} \bibnamefont{Qi}} \bibnamefont{and}
  \bibinfo{author}{\bibfnamefont{S.-C.} \bibnamefont{Zhang}},
  \bibinfo{journal}{Reviews of Modern Physics} \textbf{\bibinfo{volume}{83}},
  \bibinfo{pages}{1057} (\bibinfo{year}{2011}).

\bibitem[{\citenamefont{Qi et~al.}(2008)\citenamefont{Qi, Hughes, and
  Zhang}}]{Qi_2008}
\bibinfo{author}{\bibfnamefont{X.-L.} \bibnamefont{Qi}},
  \bibinfo{author}{\bibfnamefont{T.~L.} \bibnamefont{Hughes}},
  \bibnamefont{and} \bibinfo{author}{\bibfnamefont{S.-C.} \bibnamefont{Zhang}},
  \bibinfo{journal}{Nature Physics} \textbf{\bibinfo{volume}{4}},
  \bibinfo{pages}{273} (\bibinfo{year}{2008}).

\bibitem[{\citenamefont{Mahfouzi et~al.}(2010)\citenamefont{Mahfouzi,
  Nikoli{\'c}, Chen, and Chang}}]{mahfouzi2010}
\bibinfo{author}{\bibfnamefont{F.}~\bibnamefont{Mahfouzi}},
  \bibinfo{author}{\bibfnamefont{B.~K.} \bibnamefont{Nikoli{\'c}}},
  \bibinfo{author}{\bibfnamefont{S.-H.} \bibnamefont{Chen}}, \bibnamefont{and}
  \bibinfo{author}{\bibfnamefont{C.-R.} \bibnamefont{Chang}},
  \bibinfo{journal}{Physical Review B} \textbf{\bibinfo{volume}{82}},
  \bibinfo{pages}{195440} (\bibinfo{year}{2010}).

\bibitem[{\citenamefont{Chen et~al.}(2010)\citenamefont{Chen, Nikoli{\'c}, and
  Chang}}]{chen2010}
\bibinfo{author}{\bibfnamefont{S.-H.} \bibnamefont{Chen}},
  \bibinfo{author}{\bibfnamefont{B.~K.} \bibnamefont{Nikoli{\'c}}},
  \bibnamefont{and} \bibinfo{author}{\bibfnamefont{C.-R.} \bibnamefont{Chang}},
  \bibinfo{journal}{Physical Review B} \textbf{\bibinfo{volume}{81}},
  \bibinfo{pages}{035428} (\bibinfo{year}{2010}).

\bibitem[{\citenamefont{Luo et~al.}(2016)\citenamefont{Luo, Deng, Chen, Sheng,
  and Xing}}]{luo2016}
\bibinfo{author}{\bibfnamefont{W.}~\bibnamefont{Luo}},
  \bibinfo{author}{\bibfnamefont{W.}~\bibnamefont{Deng}},
  \bibinfo{author}{\bibfnamefont{M.}~\bibnamefont{Chen}},
  \bibinfo{author}{\bibfnamefont{L.}~\bibnamefont{Sheng}}, \bibnamefont{and}
  \bibinfo{author}{\bibfnamefont{D.}~\bibnamefont{Xing}}, \bibinfo{journal}{EPL
  (Europhysics Letters)} \textbf{\bibinfo{volume}{113}}, \bibinfo{pages}{57008}
  (\bibinfo{year}{2016}).

\bibitem[{\citenamefont{Soleimani et~al.}(2017)\citenamefont{Soleimani, Jalili,
  Mahfouzi, and Kioussis}}]{soleimani2017}
\bibinfo{author}{\bibfnamefont{M.}~\bibnamefont{Soleimani}},
  \bibinfo{author}{\bibfnamefont{S.}~\bibnamefont{Jalili}},
  \bibinfo{author}{\bibfnamefont{F.}~\bibnamefont{Mahfouzi}}, \bibnamefont{and}
  \bibinfo{author}{\bibfnamefont{N.}~\bibnamefont{Kioussis}},
  \bibinfo{journal}{EPL (Europhysics Letters)} \textbf{\bibinfo{volume}{117}},
  \bibinfo{pages}{37001} (\bibinfo{year}{2017}).

\bibitem[{\citenamefont{Meng et~al.}(2014)\citenamefont{Meng, Vishveshwara, and
  Hughes}}]{meng2014}
\bibinfo{author}{\bibfnamefont{Q.}~\bibnamefont{Meng}},
  \bibinfo{author}{\bibfnamefont{S.}~\bibnamefont{Vishveshwara}},
  \bibnamefont{and} \bibinfo{author}{\bibfnamefont{T.~L.}
  \bibnamefont{Hughes}}, \bibinfo{journal}{Phys. Rev. B}
  \textbf{\bibinfo{volume}{90}}, \bibinfo{pages}{205403}
  (\bibinfo{year}{2014}).

\bibitem[{\citenamefont{D{\'o}ra et~al.}(2012)\citenamefont{D{\'o}ra, Cayssol,
  Simon, and Moessner}}]{dora2012}
\bibinfo{author}{\bibfnamefont{B.}~\bibnamefont{D{\'o}ra}},
  \bibinfo{author}{\bibfnamefont{J.}~\bibnamefont{Cayssol}},
  \bibinfo{author}{\bibfnamefont{F.}~\bibnamefont{Simon}}, \bibnamefont{and}
  \bibinfo{author}{\bibfnamefont{R.}~\bibnamefont{Moessner}},
  \bibinfo{journal}{Physical review letters} \textbf{\bibinfo{volume}{108}},
  \bibinfo{pages}{056602} (\bibinfo{year}{2012}).

\bibitem[{\citenamefont{Vajna et~al.}(2016)\citenamefont{Vajna, Horovitz,
  D{\'o}ra, and Zar{\'a}nd}}]{vajna2016}
\bibinfo{author}{\bibfnamefont{S.}~\bibnamefont{Vajna}},
  \bibinfo{author}{\bibfnamefont{B.}~\bibnamefont{Horovitz}},
  \bibinfo{author}{\bibfnamefont{B.}~\bibnamefont{D{\'o}ra}}, \bibnamefont{and}
  \bibinfo{author}{\bibfnamefont{G.}~\bibnamefont{Zar{\'a}nd}},
  \bibinfo{journal}{Physical Review B} \textbf{\bibinfo{volume}{94}},
  \bibinfo{pages}{115145} (\bibinfo{year}{2016}).

\bibitem[{\citenamefont{Mellnik
  et~al.}(2014{\natexlab{a}})\citenamefont{Mellnik, Lee, Richardella, Grab,
  Mintun, Fischer, Vaezi, Manchon, Kim, Samarth et~al.}}]{mellnik2014}
\bibinfo{author}{\bibfnamefont{A.}~\bibnamefont{Mellnik}},
  \bibinfo{author}{\bibfnamefont{J.}~\bibnamefont{Lee}},
  \bibinfo{author}{\bibfnamefont{A.}~\bibnamefont{Richardella}},
  \bibinfo{author}{\bibfnamefont{J.}~\bibnamefont{Grab}},
  \bibinfo{author}{\bibfnamefont{P.}~\bibnamefont{Mintun}},
  \bibinfo{author}{\bibfnamefont{M.~H.} \bibnamefont{Fischer}},
  \bibinfo{author}{\bibfnamefont{A.}~\bibnamefont{Vaezi}},
  \bibinfo{author}{\bibfnamefont{A.}~\bibnamefont{Manchon}},
  \bibinfo{author}{\bibfnamefont{E.-A.} \bibnamefont{Kim}},
  \bibinfo{author}{\bibfnamefont{N.}~\bibnamefont{Samarth}},
  \bibnamefont{et~al.}, \bibinfo{journal}{Nature}
  \textbf{\bibinfo{volume}{511}}, \bibinfo{pages}{449}
  (\bibinfo{year}{2014}{\natexlab{a}}).

\bibitem[{\citenamefont{Shiomi et~al.}(2014)\citenamefont{Shiomi, Nomura,
  Kajiwara, Eto, Novak, Segawa, Ando, and Saitoh}}]{Shiomi2014}
\bibinfo{author}{\bibfnamefont{Y.}~\bibnamefont{Shiomi}},
  \bibinfo{author}{\bibfnamefont{K.}~\bibnamefont{Nomura}},
  \bibinfo{author}{\bibfnamefont{Y.}~\bibnamefont{Kajiwara}},
  \bibinfo{author}{\bibfnamefont{K.}~\bibnamefont{Eto}},
  \bibinfo{author}{\bibfnamefont{M.}~\bibnamefont{Novak}},
  \bibinfo{author}{\bibfnamefont{K.}~\bibnamefont{Segawa}},
  \bibinfo{author}{\bibfnamefont{Y.}~\bibnamefont{Ando}}, \bibnamefont{and}
  \bibinfo{author}{\bibfnamefont{E.}~\bibnamefont{Saitoh}},
  \bibinfo{journal}{Phys. Rev. Lett.} \textbf{\bibinfo{volume}{113}},
  \bibinfo{pages}{196601} (\bibinfo{year}{2014}).

\bibitem[{\citenamefont{Fan et~al.}(2014)\citenamefont{Fan, Upadhyaya, Kou,
  Lang, Takei, Wang, Tang, He, Chang, Montazeri et~al.}}]{fan2014}
\bibinfo{author}{\bibfnamefont{Y.}~\bibnamefont{Fan}},
  \bibinfo{author}{\bibfnamefont{P.}~\bibnamefont{Upadhyaya}},
  \bibinfo{author}{\bibfnamefont{X.}~\bibnamefont{Kou}},
  \bibinfo{author}{\bibfnamefont{M.}~\bibnamefont{Lang}},
  \bibinfo{author}{\bibfnamefont{S.}~\bibnamefont{Takei}},
  \bibinfo{author}{\bibfnamefont{Z.}~\bibnamefont{Wang}},
  \bibinfo{author}{\bibfnamefont{J.}~\bibnamefont{Tang}},
  \bibinfo{author}{\bibfnamefont{L.}~\bibnamefont{He}},
  \bibinfo{author}{\bibfnamefont{L.-T.} \bibnamefont{Chang}},
  \bibinfo{author}{\bibfnamefont{M.}~\bibnamefont{Montazeri}},
  \bibnamefont{et~al.}, \bibinfo{journal}{Nature materials}
  \textbf{\bibinfo{volume}{13}}, \bibinfo{pages}{699} (\bibinfo{year}{2014}).

\bibitem[{\citenamefont{Kondou et~al.}(2016)\citenamefont{Kondou, Yoshimi,
  Tsukazaki, Fukuma, Matsuno, Takahashi, Kawasaki, Tokura, and
  Otani}}]{kondou2016}
\bibinfo{author}{\bibfnamefont{K.}~\bibnamefont{Kondou}},
  \bibinfo{author}{\bibfnamefont{R.}~\bibnamefont{Yoshimi}},
  \bibinfo{author}{\bibfnamefont{A.}~\bibnamefont{Tsukazaki}},
  \bibinfo{author}{\bibfnamefont{Y.}~\bibnamefont{Fukuma}},
  \bibinfo{author}{\bibfnamefont{J.}~\bibnamefont{Matsuno}},
  \bibinfo{author}{\bibfnamefont{K.}~\bibnamefont{Takahashi}},
  \bibinfo{author}{\bibfnamefont{M.}~\bibnamefont{Kawasaki}},
  \bibinfo{author}{\bibfnamefont{Y.}~\bibnamefont{Tokura}}, \bibnamefont{and}
  \bibinfo{author}{\bibfnamefont{Y.}~\bibnamefont{Otani}},
  \bibinfo{journal}{Nature Physics} \textbf{\bibinfo{volume}{12}},
  \bibinfo{pages}{1027} (\bibinfo{year}{2016}).

\bibitem[{\citenamefont{Jamali et~al.}(2015)\citenamefont{Jamali, Lee, Jeong,
  Mahfouzi, Lv, Zhao, Nikolic, Mkhoyan, Samarth, and Wang}}]{jamali2015}
\bibinfo{author}{\bibfnamefont{M.}~\bibnamefont{Jamali}},
  \bibinfo{author}{\bibfnamefont{J.~S.} \bibnamefont{Lee}},
  \bibinfo{author}{\bibfnamefont{J.~S.} \bibnamefont{Jeong}},
  \bibinfo{author}{\bibfnamefont{F.}~\bibnamefont{Mahfouzi}},
  \bibinfo{author}{\bibfnamefont{Y.}~\bibnamefont{Lv}},
  \bibinfo{author}{\bibfnamefont{Z.}~\bibnamefont{Zhao}},
  \bibinfo{author}{\bibfnamefont{B.~K.} \bibnamefont{Nikolic}},
  \bibinfo{author}{\bibfnamefont{K.~A.} \bibnamefont{Mkhoyan}},
  \bibinfo{author}{\bibfnamefont{N.}~\bibnamefont{Samarth}}, \bibnamefont{and}
  \bibinfo{author}{\bibfnamefont{J.-P.} \bibnamefont{Wang}},
  \bibinfo{journal}{Nano letters} \textbf{\bibinfo{volume}{15}},
  \bibinfo{pages}{7126} (\bibinfo{year}{2015}).

\bibitem[{\citenamefont{Wang et~al.}(2012)\citenamefont{Wang, Sun, Chen,
  Franchini, Xu, Weng, Dai, and Fang}}]{wang2012_Na3Bi}
\bibinfo{author}{\bibfnamefont{Z.}~\bibnamefont{Wang}},
  \bibinfo{author}{\bibfnamefont{Y.}~\bibnamefont{Sun}},
  \bibinfo{author}{\bibfnamefont{X.-Q.} \bibnamefont{Chen}},
  \bibinfo{author}{\bibfnamefont{C.}~\bibnamefont{Franchini}},
  \bibinfo{author}{\bibfnamefont{G.}~\bibnamefont{Xu}},
  \bibinfo{author}{\bibfnamefont{H.}~\bibnamefont{Weng}},
  \bibinfo{author}{\bibfnamefont{X.}~\bibnamefont{Dai}}, \bibnamefont{and}
  \bibinfo{author}{\bibfnamefont{Z.}~\bibnamefont{Fang}},
  \bibinfo{journal}{Physical Review B} \textbf{\bibinfo{volume}{85}},
  \bibinfo{pages}{195320} (\bibinfo{year}{2012}).

\bibitem[{\citenamefont{Morimoto and Furusaki}(2014)}]{morimoto2014}
\bibinfo{author}{\bibfnamefont{T.}~\bibnamefont{Morimoto}} \bibnamefont{and}
  \bibinfo{author}{\bibfnamefont{A.}~\bibnamefont{Furusaki}},
  \bibinfo{journal}{Phys. Rev. B} \textbf{\bibinfo{volume}{89}},
  \bibinfo{pages}{235127} (\bibinfo{year}{2014}).

\bibitem[{\citenamefont{Yang and Nagaosa}(2014)}]{yang2014}
\bibinfo{author}{\bibfnamefont{B.-J.} \bibnamefont{Yang}} \bibnamefont{and}
  \bibinfo{author}{\bibfnamefont{N.}~\bibnamefont{Nagaosa}},
  \bibinfo{journal}{Nature communications} \textbf{\bibinfo{volume}{5}},
  \bibinfo{pages}{4898} (\bibinfo{year}{2014}).

\bibitem[{\citenamefont{Liu et~al.}(2014{\natexlab{a}})\citenamefont{Liu, Zhou,
  Zhang, Wang, Weng, Prabhakaran, Mo, Shen, Fang, Dai et~al.}}]{ZKLiu2014Na3Bi}
\bibinfo{author}{\bibfnamefont{Z.}~\bibnamefont{Liu}},
  \bibinfo{author}{\bibfnamefont{B.}~\bibnamefont{Zhou}},
  \bibinfo{author}{\bibfnamefont{Y.}~\bibnamefont{Zhang}},
  \bibinfo{author}{\bibfnamefont{Z.}~\bibnamefont{Wang}},
  \bibinfo{author}{\bibfnamefont{H.}~\bibnamefont{Weng}},
  \bibinfo{author}{\bibfnamefont{D.}~\bibnamefont{Prabhakaran}},
  \bibinfo{author}{\bibfnamefont{S.-K.} \bibnamefont{Mo}},
  \bibinfo{author}{\bibfnamefont{Z.}~\bibnamefont{Shen}},
  \bibinfo{author}{\bibfnamefont{Z.}~\bibnamefont{Fang}},
  \bibinfo{author}{\bibfnamefont{X.}~\bibnamefont{Dai}}, \bibnamefont{et~al.},
  \bibinfo{journal}{Science} \textbf{\bibinfo{volume}{343}},
  \bibinfo{pages}{864} (\bibinfo{year}{2014}{\natexlab{a}}).

\bibitem[{\citenamefont{Wang et~al.}(2013)\citenamefont{Wang, Weng, Wu, Dai,
  and Fang}}]{wang2013_Cd3As2}
\bibinfo{author}{\bibfnamefont{Z.}~\bibnamefont{Wang}},
  \bibinfo{author}{\bibfnamefont{H.}~\bibnamefont{Weng}},
  \bibinfo{author}{\bibfnamefont{Q.}~\bibnamefont{Wu}},
  \bibinfo{author}{\bibfnamefont{X.}~\bibnamefont{Dai}}, \bibnamefont{and}
  \bibinfo{author}{\bibfnamefont{Z.}~\bibnamefont{Fang}},
  \bibinfo{journal}{Physical Review B} \textbf{\bibinfo{volume}{88}},
  \bibinfo{pages}{125427} (\bibinfo{year}{2013}).

\bibitem[{\citenamefont{Liu et~al.}(2014{\natexlab{b}})\citenamefont{Liu,
  Jiang, Zhou, Wang, Zhang, Weng, Prabhakaran, Mo, Peng, Dudin
  et~al.}}]{ZKLiu2014Cd3As2}
\bibinfo{author}{\bibfnamefont{Z.}~\bibnamefont{Liu}},
  \bibinfo{author}{\bibfnamefont{J.}~\bibnamefont{Jiang}},
  \bibinfo{author}{\bibfnamefont{B.}~\bibnamefont{Zhou}},
  \bibinfo{author}{\bibfnamefont{Z.}~\bibnamefont{Wang}},
  \bibinfo{author}{\bibfnamefont{Y.}~\bibnamefont{Zhang}},
  \bibinfo{author}{\bibfnamefont{H.}~\bibnamefont{Weng}},
  \bibinfo{author}{\bibfnamefont{D.}~\bibnamefont{Prabhakaran}},
  \bibinfo{author}{\bibfnamefont{S.}~\bibnamefont{Mo}},
  \bibinfo{author}{\bibfnamefont{H.}~\bibnamefont{Peng}},
  \bibinfo{author}{\bibfnamefont{P.}~\bibnamefont{Dudin}},
  \bibnamefont{et~al.}, \bibinfo{journal}{Nature materials}
  \textbf{\bibinfo{volume}{13}}, \bibinfo{pages}{677}
  (\bibinfo{year}{2014}{\natexlab{b}}).

\bibitem[{\citenamefont{Neupane et~al.}(2014)\citenamefont{Neupane, Xu, Sankar,
  Alidoust, Bian, Liu, Belopolski, Chang, Jeng, Lin et~al.}}]{Neupane2014}
\bibinfo{author}{\bibfnamefont{M.}~\bibnamefont{Neupane}},
  \bibinfo{author}{\bibfnamefont{S.-Y.} \bibnamefont{Xu}},
  \bibinfo{author}{\bibfnamefont{R.}~\bibnamefont{Sankar}},
  \bibinfo{author}{\bibfnamefont{N.}~\bibnamefont{Alidoust}},
  \bibinfo{author}{\bibfnamefont{G.}~\bibnamefont{Bian}},
  \bibinfo{author}{\bibfnamefont{C.}~\bibnamefont{Liu}},
  \bibinfo{author}{\bibfnamefont{I.}~\bibnamefont{Belopolski}},
  \bibinfo{author}{\bibfnamefont{T.-R.} \bibnamefont{Chang}},
  \bibinfo{author}{\bibfnamefont{H.-T.} \bibnamefont{Jeng}},
  \bibinfo{author}{\bibfnamefont{H.}~\bibnamefont{Lin}}, \bibnamefont{et~al.},
  \bibinfo{journal}{Nature Communications} \textbf{\bibinfo{volume}{5}}
  (\bibinfo{year}{2014}).

\bibitem[{\citenamefont{Borisenko et~al.}(2014)\citenamefont{Borisenko, Gibson,
  Evtushinsky, Zabolotnyy, B\"uchner, and Cava}}]{borisenko2014}
\bibinfo{author}{\bibfnamefont{S.}~\bibnamefont{Borisenko}},
  \bibinfo{author}{\bibfnamefont{Q.}~\bibnamefont{Gibson}},
  \bibinfo{author}{\bibfnamefont{D.}~\bibnamefont{Evtushinsky}},
  \bibinfo{author}{\bibfnamefont{V.}~\bibnamefont{Zabolotnyy}},
  \bibinfo{author}{\bibfnamefont{B.}~\bibnamefont{B\"uchner}},
  \bibnamefont{and} \bibinfo{author}{\bibfnamefont{R.~J.} \bibnamefont{Cava}},
  \bibinfo{journal}{Phys. Rev. Lett.} \textbf{\bibinfo{volume}{113}},
  \bibinfo{pages}{027603} (\bibinfo{year}{2014}).

\bibitem[{\citenamefont{Fradkin}(1986)}]{fradkin1986critical}
\bibinfo{author}{\bibfnamefont{E.}~\bibnamefont{Fradkin}},
  \bibinfo{journal}{Physical Review B} \textbf{\bibinfo{volume}{33}},
  \bibinfo{pages}{3257} (\bibinfo{year}{1986}).

\bibitem[{\citenamefont{Ominato and Koshino}(2014)}]{ominato2014quantum}
\bibinfo{author}{\bibfnamefont{Y.}~\bibnamefont{Ominato}} \bibnamefont{and}
  \bibinfo{author}{\bibfnamefont{M.}~\bibnamefont{Koshino}},
  \bibinfo{journal}{Physical Review B} \textbf{\bibinfo{volume}{89}},
  \bibinfo{pages}{054202} (\bibinfo{year}{2014}).

\bibitem[{\citenamefont{Burkov and Kim}(2016)}]{burkov2016}
\bibinfo{author}{\bibfnamefont{A.~A.} \bibnamefont{Burkov}} \bibnamefont{and}
  \bibinfo{author}{\bibfnamefont{Y.~B.} \bibnamefont{Kim}},
  \bibinfo{journal}{Phys. Rev. Lett.} \textbf{\bibinfo{volume}{117}},
  \bibinfo{pages}{136602} (\bibinfo{year}{2016}).

\bibitem[{\citenamefont{Suzuki}(1994)}]{suzuki1994}
\bibinfo{author}{\bibfnamefont{M.}~\bibnamefont{Suzuki}},
  \bibinfo{journal}{Commun. Math. Phys.} \textbf{\bibinfo{volume}{163}},
  \bibinfo{pages}{491} (\bibinfo{year}{1994}).

\bibitem[{\citenamefont{Nakanishi et~al.}(1997)\citenamefont{Nakanishi,
  Ohtsuki, and Kawarabayashi}}]{nakanishi1997}
\bibinfo{author}{\bibfnamefont{T.}~\bibnamefont{Nakanishi}},
  \bibinfo{author}{\bibfnamefont{T.}~\bibnamefont{Ohtsuki}}, \bibnamefont{and}
  \bibinfo{author}{\bibfnamefont{T.}~\bibnamefont{Kawarabayashi}},
  \bibinfo{journal}{J. Phys. Soc. Jpn.} \textbf{\bibinfo{volume}{66}},
  \bibinfo{pages}{949} (\bibinfo{year}{1997}).

\bibitem[{\citenamefont{Chang et~al.}(2017)\citenamefont{Chang, Xu, Sanchez,
  Tsai, Huang, Chang, Hsu, Bian, Belopolski, Yu et~al.}}]{Chang_TDSM}
\bibinfo{author}{\bibfnamefont{T.-R.} \bibnamefont{Chang}},
  \bibinfo{author}{\bibfnamefont{S.-Y.} \bibnamefont{Xu}},
  \bibinfo{author}{\bibfnamefont{D.~S.} \bibnamefont{Sanchez}},
  \bibinfo{author}{\bibfnamefont{W.-F.} \bibnamefont{Tsai}},
  \bibinfo{author}{\bibfnamefont{S.-M.} \bibnamefont{Huang}},
  \bibinfo{author}{\bibfnamefont{G.}~\bibnamefont{Chang}},
  \bibinfo{author}{\bibfnamefont{C.-H.} \bibnamefont{Hsu}},
  \bibinfo{author}{\bibfnamefont{G.}~\bibnamefont{Bian}},
  \bibinfo{author}{\bibfnamefont{I.}~\bibnamefont{Belopolski}},
  \bibinfo{author}{\bibfnamefont{Z.-M.} \bibnamefont{Yu}},
  \bibnamefont{et~al.}, \bibinfo{journal}{Phys. Rev. Lett.}
  \textbf{\bibinfo{volume}{119}}, \bibinfo{pages}{026404}
  (\bibinfo{year}{2017}).

\bibitem[{\citenamefont{Ganichev et~al.}(2002)\citenamefont{Ganichev, Ivchenko,
  Bel'Kov, Tarasenko, Sollinger, Weiss, Wegscheider, and
  Prettl}}]{Ganichev_Nature2002}
\bibinfo{author}{\bibfnamefont{S.}~\bibnamefont{Ganichev}},
  \bibinfo{author}{\bibfnamefont{E.}~\bibnamefont{Ivchenko}},
  \bibinfo{author}{\bibfnamefont{V.}~\bibnamefont{Bel'Kov}},
  \bibinfo{author}{\bibfnamefont{S.}~\bibnamefont{Tarasenko}},
  \bibinfo{author}{\bibfnamefont{M.}~\bibnamefont{Sollinger}},
  \bibinfo{author}{\bibfnamefont{D.}~\bibnamefont{Weiss}},
  \bibinfo{author}{\bibfnamefont{W.}~\bibnamefont{Wegscheider}},
  \bibnamefont{and} \bibinfo{author}{\bibfnamefont{W.}~\bibnamefont{Prettl}},
  \bibinfo{journal}{Nature} \textbf{\bibinfo{volume}{417}},
  \bibinfo{pages}{153} (\bibinfo{year}{2002}).

\bibitem[{\citenamefont{Shen et~al.}(2014{\natexlab{b}})\citenamefont{Shen,
  Vignale, and Raimondi}}]{Shen_PRL2014}
\bibinfo{author}{\bibfnamefont{K.}~\bibnamefont{Shen}},
  \bibinfo{author}{\bibfnamefont{G.}~\bibnamefont{Vignale}}, \bibnamefont{and}
  \bibinfo{author}{\bibfnamefont{R.}~\bibnamefont{Raimondi}},
  \bibinfo{journal}{Phys. Rev. Lett.} \textbf{\bibinfo{volume}{112}},
  \bibinfo{pages}{096601} (\bibinfo{year}{2014}{\natexlab{b}}).

\bibitem[{\citenamefont{Chen et~al.}(2015)\citenamefont{Chen, Song, Jiang, Sun,
  Wang, and Xie}}]{Chen_PRL2015}
\bibinfo{author}{\bibfnamefont{C.-Z.} \bibnamefont{Chen}},
  \bibinfo{author}{\bibfnamefont{J.}~\bibnamefont{Song}},
  \bibinfo{author}{\bibfnamefont{H.}~\bibnamefont{Jiang}},
  \bibinfo{author}{\bibfnamefont{Q.-f.} \bibnamefont{Sun}},
  \bibinfo{author}{\bibfnamefont{Z.}~\bibnamefont{Wang}}, \bibnamefont{and}
  \bibinfo{author}{\bibfnamefont{X.~C.} \bibnamefont{Xie}},
  \bibinfo{journal}{Phys. Rev. Lett.} \textbf{\bibinfo{volume}{115}},
  \bibinfo{pages}{246603} (\bibinfo{year}{2015}).

\bibitem[{\citenamefont{Shapourian and Hughes}(2016)}]{Shapourian_PRB2016}
\bibinfo{author}{\bibfnamefont{H.}~\bibnamefont{Shapourian}} \bibnamefont{and}
  \bibinfo{author}{\bibfnamefont{T.~L.} \bibnamefont{Hughes}},
  \bibinfo{journal}{Phys. Rev. B} \textbf{\bibinfo{volume}{93}},
  \bibinfo{pages}{075108} (\bibinfo{year}{2016}).

\bibitem[{\citenamefont{Mellnik
  et~al.}(2014{\natexlab{b}})\citenamefont{Mellnik, Lee, Richardella, Grab,
  Mintun, Fischer, Vaezi, Manchon, Kim, Samarth et~al.}}]{Mellnik_2014nature}
\bibinfo{author}{\bibfnamefont{A.}~\bibnamefont{Mellnik}},
  \bibinfo{author}{\bibfnamefont{J.}~\bibnamefont{Lee}},
  \bibinfo{author}{\bibfnamefont{A.}~\bibnamefont{Richardella}},
  \bibinfo{author}{\bibfnamefont{J.}~\bibnamefont{Grab}},
  \bibinfo{author}{\bibfnamefont{P.}~\bibnamefont{Mintun}},
  \bibinfo{author}{\bibfnamefont{M.~H.} \bibnamefont{Fischer}},
  \bibinfo{author}{\bibfnamefont{A.}~\bibnamefont{Vaezi}},
  \bibinfo{author}{\bibfnamefont{A.}~\bibnamefont{Manchon}},
  \bibinfo{author}{\bibfnamefont{E.-A.} \bibnamefont{Kim}},
  \bibinfo{author}{\bibfnamefont{N.}~\bibnamefont{Samarth}},
  \bibnamefont{et~al.}, \bibinfo{journal}{Nature}
  \textbf{\bibinfo{volume}{511}}, \bibinfo{pages}{449}
  (\bibinfo{year}{2014}{\natexlab{b}}).

\bibitem[{\citenamefont{Liu et~al.}(2012)\citenamefont{Liu, Pai, Li, Tseng,
  Ralph, and Buhrman}}]{Liu_Science2012}
\bibinfo{author}{\bibfnamefont{L.}~\bibnamefont{Liu}},
  \bibinfo{author}{\bibfnamefont{C.-F.} \bibnamefont{Pai}},
  \bibinfo{author}{\bibfnamefont{Y.}~\bibnamefont{Li}},
  \bibinfo{author}{\bibfnamefont{H.}~\bibnamefont{Tseng}},
  \bibinfo{author}{\bibfnamefont{D.}~\bibnamefont{Ralph}}, \bibnamefont{and}
  \bibinfo{author}{\bibfnamefont{R.}~\bibnamefont{Buhrman}},
  \bibinfo{journal}{Science} \textbf{\bibinfo{volume}{336}},
  \bibinfo{pages}{555} (\bibinfo{year}{2012}).

\bibitem[{\citenamefont{Qiao et~al.}(2018)\citenamefont{Qiao, Zhou, Yuan, and
  Zhao}}]{Qiao_PRB2018}
\bibinfo{author}{\bibfnamefont{J.}~\bibnamefont{Qiao}},
  \bibinfo{author}{\bibfnamefont{J.}~\bibnamefont{Zhou}},
  \bibinfo{author}{\bibfnamefont{Z.}~\bibnamefont{Yuan}}, \bibnamefont{and}
  \bibinfo{author}{\bibfnamefont{W.}~\bibnamefont{Zhao}},
  \bibinfo{journal}{Phys. Rev. B} \textbf{\bibinfo{volume}{98}},
  \bibinfo{pages}{214402} (\bibinfo{year}{2018}).

\end{thebibliography}

\section*{Appendix}
To analyze the quantum transport phenomena using the tight-binding model
of the topological Dirac semimetals defined in Eqs.~(\ref{eq:TDSM}) and (\ref{eq:exchange}), we
perform the real-time evolution of the wavefunctions. Here, we present
the outline of the method. First, we decompose the Hamiltonian into
those for the odd sites and the even sites for 
$\nu=x,y,z$ directions as follows:
\begin{align}
{H}_{\rm TDSM}&={H}_{\nu,e}+{H}_{\nu,o}, \notag
\end{align}
where  ${H}_{\nu,e}$ (${H}_{\nu,o}$)
denotes the hopping process in which the origin is the even (odd) site.
Using the fourth-order Suzuki-Trotter decomposition~\cite{suzuki1994,nakanishi1997}, we decompose the
time-evolution operator ${U}$ as follows:
\begin{align}
&U(t+\Delta t,t)=
S(-{\rm i}\Delta tp,t+(1-p/2)\Delta t) \notag \\
&\times S(-{\rm i}\Delta t(1-2p),t+p\Delta t/2) \notag \\
&\times S(-{\rm i}\Delta tp,t+p\Delta t/2)
+O(\Delta t^{5}), 
\label{eq:ST}
\end{align}
where $p=(2-2^{1/3})^{-1}$.
Here, $S$ is defined as
\begin{align}
S(\eta,t)&=S_{0}(\eta,t)e^{\eta{H}_{\rm diag}}S_{1}(\eta,t),\notag \\
S_{0}(\eta,t)&=e^{\eta{H}_{x,e}/2}e^{\eta{H}_{x,o}/2}e^{\eta{H}_{y,e}(t)/2} \notag \\
&\times e^{\eta{H}_{y,o}(t)/2}e^{\eta{H}_{z,e}/2}e^{\eta{H}_{z,o}/2}\notag \\
S_{1}(\eta,t)&=e^{\eta{H}_{z,o}/2}e^{\eta{H}_{z,e}/2}e^{\eta{H}_{y,o}(t)/2} \notag \\
&\times e^{\eta{H}_{y,e}(t)/2}e^{\eta{H}_{x,o}/2}e^{\eta{H}_{x,e}/2}, \notag
\end{align}
where $H_{\rm diag}$ denotes the diagonal part of the
Hamiltonian such as the chemical potential.
Using Eq.~(\ref{eq:ST}), we perform the real-time evolution
of the wave function.

\end{document}